%% file: main_recsys_2025.tex
\documentclass[sigconf]{acmart}
\usepackage[utf8]{inputenc}
\usepackage[T1]{fontenc}
\usepackage{amsfonts}
\usepackage{mathtools}
\usepackage{graphicx}
\usepackage{hyperref}
\PassOptionsToPackage{hyphens}{url}\usepackage{hyperref}
\usepackage{natbib}
\usepackage{setspace}
\usepackage{enumitem}
\usepackage[linesnumbered,titlenotnumbered,ruled,vlined]{algorithm2e}
\usepackage{array}
\usepackage{float}
\usepackage[format=plain,
            labelfont=it,
            textfont=it,
            skip=2.5pt]{caption}
\usepackage{natbib}
\usepackage{subcaption}
\usepackage{balance}

\hypersetup{colorlinks,linkcolor={blue},citecolor={blue},urlcolor={red}} 

\newenvironment{proofS}{%
  \proof}{\endproof}

\theoremstyle{plain}
\newtheorem{theorem}{Theorem}
\newtheorem*{remark}{Remark}

\theoremstyle{definition}

\theoremstyle{remark}



\copyrightyear{2025}
\acmYear{2025}
\setcopyright{none}
\setcctype{by}
\acmConference[RecSys '25]{Proceedings of the Nineteenth ACM Conference on Recommender Systems}{September 22--26, 2025}{Prague, Czech Republic}
\acmBooktitle{Proceedings of the Nineteenth ACM Conference on Recommender Systems (RecSys '25), September 22--26, 2025, Prague, Czech Republic}\acmDOI{10.1145/3705328.3748063}
\acmISBN{979-8-4007-1364-4/2025/09}

\author{Md Sanzeed Anwar}
\affiliation{%
  \institution{University of Michigan}
  \city{Ann Arbor}
  \state{MI}
  \country{USA}}
  \email{sanzeed@umich.edu}

\author{Paramveer S. Dhillon}
\affiliation{%
  \institution{University of Michigan}
  \city{Ann Arbor}
  \state{MI}
  \country{USA}}
  \email{dhillonp@umich.edu}

\author{Grant Schoenebeck}
\affiliation{%
  \institution{University of Michigan}
  \city{Ann Arbor}
  \state{MI}
  \country{USA}}
  \email{schoeneb@umich.edu}

\begin{document}

\title[Recommendation and Temptation]{Recommendation and Temptation}

\begin{abstract}
Traditional recommender systems based on revealed preferences often fail to capture the fundamental duality in user behavior, where consumption choices are driven by both inherent value (enrichment) and instant appeal (temptation). Consequently, these systems may generate recommendations that prioritize short-term engagement over long-lasting user satisfaction. We propose a novel recommender design that explicitly models the tension between enrichment and temptation. We introduce a behavioral model that accounts for how both enrichment and temptation influence user choices, while incorporating the reality of off-platform alternatives. Building on this model, we formulate a novel recommendation objective aligned with maximizing consumed enrichment and prove the optimality of a locally greedy recommendation strategy. Finally, we present an estimation framework that leverages the distinction between explicit user feedback and implicit choice data while making minimal assumptions about off-platform options. Through comprehensive evaluation using both synthetic simulations and real-world data from the MovieLens dataset, we demonstrate that our approach consistently outperforms competitive baselines that ignore temptation dynamics either by assuming revealed preferences or recommending solely based on enrichment. Our work represents a paradigm shift toward more nuanced and user-centric recommender design, with significant implications for developing responsible AI systems that genuinely serve users' long-term interests rather than merely maximizing engagement.
\end{abstract}

\maketitle

\section{Introduction}
Consumption of online content has become a central fixture in everyday life, spanning various forms such as articles, videos, films, social media, and more. This consumption accounts for a substantial portion of people's time, significantly impacting daily routines and leisure activities. For example, according to Statista\footnote{\url{https://www.statista.com/statistics/1359403/us-time-spent-per-day-netflix-tiktok-youtube/}}, adults in the United States spend an average of 46 minutes on YouTube, 48 minutes on TikTok, and 61 minutes on Netflix \emph{every day}. Recommender systems act as the guiding force during these excursions, helping users navigate the vast sea of content more effectively.  The traditional recommendation design paradigm is rooted in behaviorism, relying on observed user behavior to predict and guide future interactions. Initially implemented using matrix factorization techniques popularized by the 2006 Netflix challenge, modern recommenders now employ sophisticated machine learning approaches, including neural networks for user-item feature matching \cite{He2017} and graph neural networks for capturing user-item interaction networks \cite{Wang2019, He2020} etc.

However, there is a growing concern that these systems may be guiding users to the wrong sets of content, therefore preventing them from getting the maximum benefit possible out of consumption. This potential shortcoming stems from two core premises of the current recommender design paradigm:

\begin{enumerate}
\item Each user has a single utility function that represents the benefit they derive from a given option.
\item Users are utility maximizers and, when presented with a set of alternatives, will invariably choose the option with the highest utility.
\end{enumerate}

These assumptions, while providing a foundation for conventional recommenders, oversimplify the complex nature of human decision-making during content consumption.

\subsection{Temptation}
In this paper, we present an overhaul of recommender design by challenging these premises. \textbf{Our first contribution is a novel behavioral model grounded in the literature on temptation and has the following three features}:

\begin{enumerate}
    \item \textbf{Enrichment:} Each item has some inherent value that the user derives from consuming it.
    
    \item \textbf{Temptation:} Each item also has some inherent appeal that is orthogonal to its enrichment or inherent value, but nonetheless influences the consumption choices made by users. 
    \textit{Note that the enrichment and temptation of an item is personalized for a given user, and can vary across users.}
    
    \item \textbf{Outside Options:} While the platform presents the user with a set of recommended items, the user has an additional outside option available to them. The user chooses their consumption from the recommended items and their available outside option according to a personalized weighted combination of enrichment and temptation.
\end{enumerate}

The intuition is that while users would benefit from consuming highly enriching content, they may instead be tempted by options with inferior enrichment but high appeal, thus failing to act in their own best interest. This violates the first premise because users effectively operate with two utility functions: one representing their true preferences or end goals, and another representing their actual choice behavior. It also violates the second premise because users make consumption decisions by optimizing their choice behavior rather than their stated preferences, appearing to contradict the assumption of utility maximization.

\vspace*{7.5pt}

The human tendency to do other than that which an individual wants is already observed by Paul in the Christian Bible, ``I do not understand my own actions. For I do not do what I want, but I do the very thing I hate" (Romans 7:15).  More recently, the research literature documents this mismatch across domains ranging from purchase behaviors \cite{Klaus1998} and appointment compliance \cite{trope2000} to smoking cessation \cite{gine2010}, savings \cite{ThalerShlomo2004}, and study concentration \cite{ariely2002}. In the context of digital media, Milkman et al. \cite{milkman2009highbrow} ran an empirical study on Quickflix and examined the orders in which DVDs were rented and returned, illustrating a difference between stated preferences (adding DVDs to the rental queue) and consumption (watch order of DVDs).

Economists have proposed several theoretical frameworks to explain this behavior. One class of models, including those by Strotz \cite{Strotz1955}, Akerlof \cite{Akerlof1991}, Laibson \cite{Laibson1997}, and O'Donoghue and Rabin \cite{Donoghue1999}, suggests that people exhibit time-inconsistent preferences, prioritizing immediate benefits in ways that contradict their previously stated preferences. A second class of models, proposed by Thaler and Shefrin \cite{Thaler1981} and others, introduces the concept of dual selves: a forward-looking self concerned with long-term benefits and a myopic self focused on instant gratification, with the latter exerting greater influence over actual decisions. A third class of models, developed by Gul and Pesendorfer \cite{Pesendorfer2001, Pesendorfer2004}, explains user behavior by treating temptation as an additional attribute of options beyond intrinsic value, showing how it influences commitment and decision-making processes.

Our model remains agnostic about the precise nature of enrichment and temptation. In our model, enrichment simply represents what users genuinely wish to optimize, whether their stated preferences, long-term benefits, or other intrinsic values. Temptation, in turn, captures users' immediate desire to choose an item, independent of its enrichment value.

\vspace*{7.5pt}

Our model implies that users' observed behaviors from engagement may not serve as reliable indicators of their true preferences (e.g. revealed preferences) or derived value. Consequently, traditional recommenders that rely on behavioral signals and optimize for engagement metrics cannot effectively optimize for users' stated desires or long-term well-being. Beginning with Ekstrand and Willemsen \cite{Ekstrand2016}, a growing body of literature has documented this fundamental limitation of engagement-based recommenders and often desiring instead to incorporate some form of explicit user feedback. Some scholars even characterize the continued reliance on revealed preferences as ``indefensible'' \cite{Morewedge2023}.

This aligns with qualitative evidence showing that users express hindsight regret when evaluating their experience on content platforms~\cite{mccrosky2021youtube}. Indeed, users on platforms like YouTube report feeling ``sucked down a rabbit hole'' and engaging in prolonged consumption of content they themselves acknowledge as mediocre~\cite{woolley2023rabbit}.

Milli et al.~\cite{milli2023engagement} demonstrated in an empirical study on Twitter (now X) that users experienced reduced negative emotions and greater satisfaction from political tweets recommended based on stated preferences (e.g., all posts from those they follow) compared to actual engagement-based recommendations. Their results not only demonstrate the divergence between engagement-based recommendations and those based on stated preferences, but also reveal that engagement-based recommendations contain more anti-social content, highlighting potential societal harms beyond individual user enrichment loss.

Furthermore, recently Kleinberg et al.~\cite{Kleinberg2022} have used a dual-self model to illustrate how users become drawn to and are unable to disengage from highly tempting content, despite recognizing its limited long-term value or enrichment potential. Our behavioral model draws inspiration from theirs; however, ours is not based on a particular characterization of users' inner selves.  While both models are concerned with how tempting items may distort platform use, their model focuses on time-on-platform whereas ours focuses on which items the user chooses to consume.  Additionally, we go beyond merely diagnosing user behavior to prescribe a recommender system that actively helps users overcome temptation.

\subsection{Optimizing Enrichment}
The literature discussed so far demonstrates the shortcoming of recommenders optimizing engagement metrics in helping users achieve their end goal, i.e. optimize their enrichment. An alternative is to simply try to reconstruct the enrichment of each item and recommend the items with the highest enrichment. However, in most realistic settings, this fails as illustrated in the following anecdote: one of the authors signed up for a Netflix subscription at the time when they shipped DVDs. The first DVD they ordered was ``Lawrence of Arabia," a nearly four hour biopic on T.~E.~Lawrence. They returned it 4 months later, unwatched.

Outside options are also crucial to consider when optimizing enrichment. In realistic scenarios, content platforms are not insulated \textemdash they exist in an ecosystem with competing platforms and activities, framed as exogenous off-platform options available to a user when making consumption choices in our model. The author did finally watch ``Lawrence of Arabia" on a 15-hour flight, where outside options were limited.

Industry leaders recognize the necessity of considering outside options in platform and recommender design. Luis von Ahn, creator of Duolingo, talks about ``How to Make Learning as Addictive as Social Media''~\cite{vonahn2023learning,raz2020recaptcha} as a necessity to combat the distraction of other apps on a user's smartphone. He acknowledges navigating a tradeoff between the educational value of Duolingo and its ability to retain users.  Similarly, Netflix CEO Reed Hastings talks about Netflix competing not just with other streamers but anything users do to ``relax and unwind, hang out and connect'' including sleep~\cite{cnbc2017netflix}.

These challenges in optimizing enrichment arise from a fundamental insight captured by our model: recommended items need not resemble consumed items. There is a crucial distinction between \textit{recommended} enrichment and \textit{consumed} enrichment. While recommending highly engaging content is problematic, simply recommending enriching items may not be much better, as echoed by industry leaders who recognize this tension. Instead, platforms must carefully consider temptation when making recommendations. Moreover, platforms must exercise caution even when making multiple recommendations without outside options, as a single highly tempting item can deter users from selecting other recommendations regardless of their enrichment value.

Understanding users' outside options is equally crucial for optimizing enrichment. Consider two contrasting scenarios: a user seeking quality content during leisure time that would otherwise be spent on low-value activities like doom scrolling, versus a student procrastinating on important exam preparation. In the first case, the platform should recommend sufficiently engaging content to retain the user while maximizing enrichment within this constraint. In the second case, where any platform content is likely less enriching than studying, maximizing enrichment requires helping the user avoid the platform altogether by recommending minimally tempting content or no recommendations at all. This illustrates how optimal recommendation strategies must dynamically adapt based on the relative enrichment value of users' available alternatives, fundamentally challenging the one-size-fits-all approach of traditional recommender systems.

\textbf{Our second contribution is to formulate a novel optimization objective for maximizing \textit{consumed} enrichment and ensuring user well-being.} In addition, we also prescribe the optimal recommendation strategy for this objective, with proof of optimality. Our model and optimization provide three key implications for design:
\begin{enumerate}
    \item Consumption choices made by a user may prove to be a poor proxy for their end goal or true desire. 
    \item Items recommended do not necessarily equal items consumed, so it is not necessarily enough to recommend highly enriching items.
    \item Recommending items can be costly. Tempting items may crowd out more enriching, but less tempting alternatives.
    \item Maximizing enrichment requires understanding how enriching the outside option is, and whether the platform should be trying to retain the user while maximizing enrichment.  
\end{enumerate}
  
We note here that our model captures the settings where a user must choose which option to consume. Therefore, our results apply more to content platforms like YouTube and Netflix, than to social network platforms like X, Bluesky, Instagram, or Facebook, where the consumption of each item is nearly instantaneous and recommendation almost certainly implies consumption.  

\subsection{Learning from Data}
Our proposed recommender algorithm requires knowledge of enrichment and temptation values for items. In practice, these are unobservable quantities that must be learned from data.  

Existing literature offers several approaches to this challenge. Agarwal et al.\cite{Agarwal2024} examine user return probability as a behavioral proxy for long-term satisfaction. Chang et al.\cite{Chang2023} employ a latent variable model where past behaviors and context generate probability distributions over hidden user intents, which then predict future behavior.  However, most prior studies combine solicited non-engagement signals with engagement metrics to infer user satisfaction. For instance, Cunningham et al.\cite{cunningham2024} propose using engagement diversity and item-level surveys. Milli et al.\cite{Milli2021} demonstrate how to use specific observed behaviors as anchor variables to derive ``value'' rather than engagement, such as leveraging a ``show me less of this'' button.  Other researchers suggest that using multiple behavioral signals rather than a single metric could provide better insights~\cite{Kleinberg2024}. Indeed, Milli et al.~\cite{Milli2023} show that up-weighting behaviors more indicative of whether users value specific content leads to improved ``value'' estimation in their model.

\textbf{Our third contribution is a basic estimation framework that identifies enrichment and temptation from a combination of behavioral data and explicit user feedback about both on-platform items and outside options.} Crucially, our framework imposes minimal, reasonable assumptions about users' outside options, which are necessary for implementing the optimal recommender. This serves to demonstrate the feasibility of our proposed recommender in practical scenarios.  

Our estimator is designed for our specific setting; however, in practice, ideas and insights from the aforementioned works could likely be combined with our model.

\section{Behavioral Model}\label{model}

We model user behavior within the confines of a single platform, considering $m$ users and $n$ items, where $m \gg n$. As previously discussed, for each user $j = 1, \ldots, m$, each item $i = 1, \ldots, n$ has two attributes:
i) \textbf{\textit{Enrichment}} $u_j(i)$ measures the inherent value the item provides to the user, and ii) \textbf{\textit{Temptation}} $v_j(i)$ measures the appeal of the item to the user.

To capture the reality that users have alternatives to platform engagement, we introduce the concept of outside options. We assume that in addition to items on the platform, each user $j$ also has $K$ outside options $o^1_j, \ldots, o^K_j$. At any time $t$, only one of these outside options $o_j(t)$, chosen according to some probability distribution, is available to them as part of their set of choices. These outside options represent a user's freedom to not engage in consumption on the platform, adding a layer of realism to our model.

In our model, consumption takes place in $T$ discrete rounds. In each round $t$, the decision-making process for user $j$ unfolds as follows:

\begin{enumerate}[leftmargin=*]
\item \textbf{Recommendation stage:} User $j$ is presented with a set $S_j(t)$ of available on-platform items. Therefore, the set of choices for user $j$ in round $t$ is given by $S_j(t) \cup \{o_j(t)\}$, encompassing both recommended items and the available outside option.

\item \textbf{Consumption stage:} User $j$ then chooses an option by maximizing their \textbf{choice score} of available items. The choice score of an item $i$ for user $j$ is defined as:

\[\lambda^{C}_j u_j(i) + (1 - \lambda^{C}_j) v_j(i)\]

Here, $\lambda^{C}_j \in [0, 1]$ is the \textbf{choice parameter} representing the user's relative preference for enrichment over temptation when choosing what to consume. For instance, $\lambda^{C}_j < 0.5$ implies that the user places more weight on instant gratification when making their choice.

Therefore, in round $t$, user $j$ selects the option $i^*(t) \in S_j(t) \cup \{o_j(t)\}$ such that:

\[
i^*(t) = \operatorname*{argmax}_{i \in S_j(t) \cup \{o_j(t)\}} \biggl(\lambda^{C}_j u_j(i) + (1 - \lambda^{C}_j) v_j(i)\biggr),
\]

\end{enumerate}
This formulation elegantly captures the user's decision-making process, taking into account both the enrichment and temptation of each item, as well as the available outside option.

To model content variety and prevent repetitive consumption, we stipulate that once user $j$ chooses to consume an item on the platform, it becomes unavailable to them for future consumption.
\paragraph{ \noindent \textbf{User feedback:}}
Our model also incorporates user feedback, an essential component for many recommender systems. Upon consuming an item $i$, user $j$ provides explicit feedback (e.g., ratings) $r_j(i)$ about the item. This feedback is defined as a function of the \textbf{feedback score}. The feedback score of an item $i$ for user $j$ is:

\[\lambda^{F}_j u_j(i) + (1 - \lambda^{F}_j)v_j(i)\]

Here, $\lambda^{F}_j \in [0, 1]$ is a \textbf{feedback parameter} that represents the extent to which the user's explicit feedback depends on enrichment versus temptation. For example, $\lambda^{F}_j > 0.5$ indicates that the user's explicit feedback for an item is determined primarily by its enrichment rather than its temptation.

Formally, the explicit feedback of user $j$ for item $i$ is given as:

\[
r_j(i) = f_{rating}\left(\lambda^{F}_j u_j(i) + (1 - \lambda^{F}_j)v_j(i)\right).
\]

where $f_{rating}$ is monotone and non-decreasing. Notice that our conception of rating data need not directly reveal a user's enrichment and may also be influenced by the temptation of items.

An important assumption in our model is that $\lambda^F_j > \lambda^C_j \hspace*{2mm}\forall j=1,\ldots,m$. This means that users tend to place more weight on enrichment when providing feedback than when choosing what to consume. This assumption is grounded in the observation that rating an item typically involves more deliberate reflection from a user compared to the often more impulsive act of choosing what to consume. This aligns with psychological theories suggesting that people are more likely to consider the inherent value when engaged in reflective thought processes.

\section{Recommender Design}\label{recommender_design}
We aim to design a recommendation system that ensures as much enrichment from consumption as possible. Consequently, we adopt the following objective function for our recommendation design:

\[\max_{\{S_j(1),\ldots, S_j(T)\}}\left\{\mathbb{E}\left[\sum_{t = 1}^{T} u_j\left(i^*(t)\right)\right]\right\}\]

It's crucial to understand that this objective does not merely maximize the enrichment of recommended items, but rather aims to optimize the expected enrichment users receive from consumption. This approach is grounded in the recognition that recommending high-enrichment items does not ensure user engagement. The user's choice is determined by the item's choice score, which is a weighted combination of both its enrichment and temptation. Consequently, items with high enrichment but low temptation might be overlooked in favor of items with moderate enrichment but high temptation.

The question that naturally follows is: which recommendation strategy is optimal given this objective? The answer, perhaps surprisingly, is a ``locally greedy'' strategy. It chooses what to recommend by maximizing the expected enrichment in a single round from an item. The intuition behind this strategy is twofold:

\begin{enumerate}[leftmargin=*]
\item recommending items with high temptation but moderate/low enrichment might inadvertently crowd out on/off-platform choices with higher enrichment; \item solely recommending  high-enrichment items may yield a selection of items that are insufficiently attractive to overcome the enticement of off-platform items with lower enrichment but higher temptation. \end{enumerate}  

Thus, when considering an item the recommender will balance the additional expected enrichment if the user selects that item and the likelihood that item will be selected by the user.

We prove the optimality of this recommendation strategy in a perfect information scenario: where the platform knows the enrichment and temptation of both on-platform items and off-platform outside options, the probability distribution that determines the availability of outside options for each user, and the user-specific parameters. The performance of this recommendation strategy in practice would hence depend on how well the platform can emulate perfect information from available historical data.

We denote by $u_j(i|o)$ the enrichment user $j$ gets from consumption in a single round when the available options are item $i$ and outside option $o$. Specifically,
\[u_j(i|o) = \begin{cases} 
u_j(i) & \text{if } \hspace*{2mm}\begin{aligned}
    &\lambda^{C}_j u_j(i) + (1 - \lambda^{C}_j) v_j(i)\\ &\geq \lambda^{C}_j u_j(o) + (1 - \lambda^{C}_j) v_j(o) 
\end{aligned}\\
u_j(o) & \text{otherwise}
\end{cases}\]

This formulation captures the user's decision-making process, taking into account both the enrichment and temptation of the available options. It forms the basis for our optimal recommendation strategy, which we will explore in more detail in the following theorem.

\begin{theorem}\label{main_theorem}
    When the platform has perfect knowledge of everything except the exact availability of outside options, the optimal recommendation strategy is \textbf{locally greedy}, i.e. in each round $t$, it recommends the available item that maximizes the expected enrichment a receives from consumption in a single round when the available options are the said item and the outside option.
    
    In other words, the optimal strategy recommends the available item $i^*$ in round $t$ such that
    \[i^* = \operatorname*{argmax}_{i}\mathbb{E}_{o_j(t)}\left[u_j\left(i|o_j(t)\right)\right]\]
\end{theorem}

\begin{proofS}
    Note that because a user's choice process is deterministic, recommending a set of items is equivalent to recommending a single item (the item with the highest choice score in the set). Thus, any recommendation strategy can be represented as a tree, where each node represents the recommended item, and we branch based on whether the recommended item was chosen or not. 
    
    We then induct on the number of rounds (the depth of the tree), $T$. The intuition for the proof here is that if a recommendation strategy $\mathcal{A}$ is optimal, it must be locally greedy after the first round because of the induction hypothesis. Then, we can either swap the item recommended by $\mathcal{A}$ in round 1 with a later item without losing enrichment, or we can replace the item recommended by $\mathcal{A}$ in round 1 with an item that provides better enrichment in expectation. A detailed proof is provided as auxiliary material. 
\end{proofS}

Results from synthetic simulations and simulations based on real-world data in a perfect information scenario, shown in figure \ref{fig:all_performance}, empirically demonstrates this optimality.

\begin{remark} 
For traditional recommenders, recommending an additional item cannot hurt: the agent may spend additional time inspecting it, but could simply not choose it. In contrast, in our model, additional recommendations can dramatically harm user enrichment. In fact, in theory, the enrichment score can be made arbitrarily worse by recommending an item with a sufficiently low enrichment so that it would be harmful if chosen, but a sufficiently high temptation so that it is indeed chosen by users.
\end{remark}

\section{Model Estimation}\label{estimation}
The optimal recommender strategy proposed in section \ref{recommender_design} requires knowledge of both on-platform items and off-platform outside options for each user, specifically their enrichment and temptation values, as well as user-specific parameters. However, in reality, platforms do not possess this comprehensive information and must instead estimate these components from available historical data. To address this challenge, we first extend our model to elucidate the determination of enrichment and temptation values for items. Subsequently, we provide a framework for estimating these critical components and generating recommendations that align with our proposed strategy. This estimation process bridges the gap between theoretical optimality and practical implementation, demonstrating that platforms can in fact leverage our recommendation strategy effectively, even in the absence of perfect information. By doing so, we not only enhance the applicability of our model but also provide valuable insights into user behavior and content characteristics.

\subsection{Enrichment and Temptation}

We adopt low-dimensional vector representations for both users and items, a common approach in recommender systems. Specifically, we assume that the enrichment and temptation yielded by item i for user j are given by:
\[
u_j(i) = \mathbf{a}^{\top}_j \mathbf{x}_i, \quad v_j(i) = \mathbf{b}^{\top}_j \mathbf{y}_i.
\]

where $\mathbf{a}_j$, $\mathbf{b}_j$, $\mathbf{x}_i$ and $\mathbf{y}_i$ are all $d$-dimensional vectors (we assume $d \ll m, n$) defined as follows:

\begin{enumerate}[leftmargin=*]

\item $\mathbf{a}_j$ represents user $j$ in the enrichment space, while $\mathbf{b}_j$ represents them in the temptation space.

\item $\mathbf{x}_i$ represents item $i$ in the enrichment space, while $\mathbf{y}_i$ represents it in the temptation space.

\end{enumerate}

We assume that each item $i$ yields some universal enrichment and temptation, represented by the first components of $\mathbf{x}_i$ and $\mathbf{y}_i$, respectively. We represent the universality of these values by setting the first components of $\mathbf{a}_j$ and $\mathbf{b}_j$ to 1 for each user $j$:

\[\mathbf{a}_{j1} = 1, \quad \mathbf{b}_{j1} = 1 \quad \forall j = 1,\ldots, m.\]

The remaining components of $\mathbf{a}_j$ and $\mathbf{b}_j$ represent user-specific idiosyncrasies. Thus, we assume that these components contain no systemic biases, i.e., $\mathbb{E}_j[\mathbf{a}_{j\ell}] = \mathbb{E}_j[\mathbf{b}_{j\ell}] = 0, \forall \ell = 2,\ldots, d$.

This formulation allows for a clear separation of the universal and user-specific aspects of enrichment and temptation, while maintaining a concise representation using vector dot products. Consequently, we have three types of model parameters that we need to estimate:
\begin{enumerate}[leftmargin=*]
    \item \textbf{User-specific parameters:} Feedback parameter $\lambda^F_j$ and choice parameter $\lambda^C_j$ for user $j$.

    \item \textbf{Item-specific parameters:} Enrichment $u_j(i)$ and temptation $v_j(i)$ of item $i$ for user $j$.

    \item \textbf{Outside option-specific parameters:} Enrichment $u_j(o)$ and choice score $\lambda^{C}_j u_j(o) + (1 - \lambda^{C}_j) v_j(o)$ for outside option $o$ and user $j$.
\end{enumerate}

\subsection{Estimation Framework}
To implement our proposed recommendation strategy, the platform must estimate all three types of model parameters mentioned above. We work with \textbf{two types of historical data}:

\begin{enumerate}[leftmargin=*]
\item \textbf{Behavioral data:} The set of items, $S_j(t)$, recommended to each user $j$ in each round $t$, along with the user's subsequent choice. Notably, we assume the platform can identify when a user opts for an outside option over recommended items, but cannot determine which specific outside option was selected.

\item \textbf{Explicit feedback:} Rating $r_{ij}$ provided by user $j$ after consuming item $i$. Crucially, we assume the platform can compute the feedback score from the provided rating, and vice versa.
\end{enumerate}

Limiting the platform's access to only these two data types presents a significant challenge for effective recommendation design. Any recommendation strategy that fails to learn about and account for the outside option-specific parameters cannot achieve optimality. Because what items a user consumes (and consequently what enrichment they receive) depends not only on the recommended items, but also on the available outside options. To address this limitation and enable the implementation of our recommendation strategy, we introduce two additional assumptions about outside option-specific parameters:

\begin{enumerate}[leftmargin=*]
    \item We posit that \textbf{the choice scores of outside options adhere to a known distributional form}. This allows the platform to estimate the distribution parameters from available data, enabling the recommendation algorithm to compute outside option choice scores without requiring detailed knowledge of each user's specific outside options. For our purposes, we assume these choice scores follow a normal distribution $\mathcal{N}(\mu, \sigma^2)$, and we only need to estimate $\mu$ and $\sigma$. However, note that our framework could easily be extended to other distributions.
    
    \item We further assume that enrichment and choice scores of outside options are independent, and that \textbf{the platform can ascertain the expected enrichment of outside options}, denoted as $\mathbb{E}_{\ell}\left[u_j(o^\ell_j)\right]$. This information can be reasonably obtained directly from users, for example, through brief surveys about their off-platform experiences.
\end{enumerate}

These assumptions, while providing crucial information, remain reasonable for adoption by real-world platforms. They offer the minimum necessary data for our proposed recommendation strategy to function effectively. Given these assumptions, we can solve for user-specific, item-specific and outside option-specific parameters. The detailed optimization process for this estimation is outlined in Algorithm \ref{alg:est_all}. \footnote{Code used for the experiments available at \href{https://github.com/Sanzeed/recommendation-and-temptation}{https://github.com/Sanzeed/recommendation-and-temptation}}

Once we have estimated these model parameters, we can plug them into the closed form for the expected enrichment user $j$ would receive from a single round of consumption if some item $i$ was recommended:
\begin{align*}
    &\mathbb{E}\left[u_j\left(i|o_j(t)\right)\right] \\= &\mathbb{P}\left[j \text{ chooses i}\right] u_j(i) + \left(1 - \mathbb{P}\left[j \text{ chooses i}\right]\right) \mathbb{E}\left[u_j(o_j(t))\right]\\
    = & \phi\left(\frac{C_j(i) - \mu}{\sigma}\right)u_j(i) + \left(1 - \phi\left(\frac{C_j(i) - \mu}{\sigma}\right)\right) \mathbb{E}\left[u_j(o_j(t))\right]
\end{align*}
where $C_j(i) = \lambda^{C}_j u_j(i) + (1 - \lambda^{C}_j) v_j(i)$ is the choice score of item $i$ for user $j$, and $\phi$ is the CDF of the standard normal distribution. Given this closed form, we can once again use the optimal greedy recommendation strategy as desired. 

\begin{algorithm}
    \SetKwInOut{Input}{Input}
    \SetKwInOut{Output}{Output}
    
    \caption{Model Estimation}
    \label{alg:est_all}
    
    \Input{Recommendation $S_j(t)$ for each $j$ in each past $t$;\\
    Chosen items $i^*_j(t)$ for each $j$ in each past;\\
    Set $R$ of explicit feedback $r_j(i)$;\\}
    \Output{Estimations $\hat{u}_j(i), \hat{v}_j(i), \hat{\lambda}^F_j, \hat{\lambda}^C_j, \mu, \sigma$ \\ $\forall j =1, \ldots, m; i = 1, \ldots, n$}
    
    \Begin{
        Estimate $\hat{\mathbf{a}}_j, \hat{\mathbf{b}}_j, \hat{\mathbf{x}}_i, \hat{\mathbf{y}}_i, \hat{\lambda}^F_j, \hat{\lambda}^C_j, \mu, \sigma$ via stochastic gradient descent with the following loss:
        \begin{itemize}
            \item Est. rating, $\hat{r}_j(i) = f_{rating}\left(\hat{\lambda}^{F}_j \hat{\mathbf{a}}^\top_j\hat{\mathbf{x}}_i + (1 - \hat{\lambda}^{F}_j)\hat{\mathbf{b}}^\top_j\hat{\mathbf{y}}_i\right)$
            \item Total loss from rating data, $L_{rating} = \sum_{\substack{j, i: r_j(i)\in R}} \left(r_j(i) - \hat{r}_j(i)\right)^2$
            \item Est. choice value, $\hat{C}_j(i) = \hat{\lambda}^{C}_j \hat{\mathbf{a}}^\top_j\hat{\mathbf{x}}_i + (1 - \hat{\lambda}^{C}_j)\hat{\mathbf{b}}^\top_j\hat{\mathbf{y}}_i$
            \item Est. outside option choice value, $\hat{C}_j(o_j(t)) \sim \mathcal{N}(\mu, \sigma)$
            \item Per user per round loss, $H(j, t)$ $=\sum_{i\in S_j(t)\cup\{o_j(t)\}}\min\left(0, \hat{C}_j(i^*(t)) - \hat{C}_j(i)\right)$
            \item Total loss from recommendation and choice data, $L_{click} = \sum_{j, t} H(j, t)$
            \vspace*{2mm}
            \item \textbf{Total loss,} $L = \alpha L_{rating} + \beta L_{click}$\\ 
            for appropriate weights $\alpha$, $\beta$ with $\alpha + \beta = 1$
        \end{itemize}

        Compute $\hat{u}_j(i) = \hat{\mathbf{a}}^\top_j\hat{\mathbf{x}}_i, \hat{v}_j(i) = \hat{\mathbf{b}}^\top_j\hat{\mathbf{y}}_i$\;
        Return $\hat{u}_j(i), \hat{v}_j(i), \hat{\lambda}^F_j, \hat{\lambda}^C_j, \mu, \sigma$ for all $j, i$\;
    }
\end{algorithm}

\section{Evidence from Simulation}\label{simulation}

Given the challenges of conducting in-situ tests with recommender systems, we turn to comprehensive agent-based simulations, a widely accepted approach in the field for empirical validation \cite{Chaney2018, Nguyen2014, fleder2009blockbuster}. Our simulation environment comprises $m = 1000$ users and $n = 250$ items, situated in $d$-dimensional ($d=3$) enrichment and temptation spaces. We generate vector representations for users and items from multivariate normal distributions, with parameters carefully chosen to induce a weak anti-correlation between the enrichment and temptation of items for any given user. 

In our simulation, all users have access to the same set of $K = 100$ outside options. We draw the enrichment and temptation space representations of these outside options from multivariate normal distributions as well. In particular, we draw the vector representations of on-platform items and off-platform outside options to simulate three distinct scenarios:

\begin{enumerate}[leftmargin=*]
    \item \textbf{Enriching on-platform items:} We are primarily interested in exploring the scenario where on-platform items have higher enrichment and lower temptation than outside options on average.
    \item \textbf{Tempting on-platform items:} As a secondary inquiry, we also explore the scenario where on-platform items have lower enrichment and higher temptation than outside options on average.
    \item \textbf{Similar on and off-platform items:} We also simulate recommendations in the scenario where the on and off-platform items have the same enrichment and temptation on average.
\end{enumerate}

For each user, we independently sample the feedback parameter $\lambda^F_j$ and choice parameter $\lambda^C_j$ from unimodal beta distributions. We set the distribution parameters such that the mode of $\lambda^F_j$ is approximately $0.75$, while the mode of $\lambda^C_j$ is approximately $0.25$. To maintain consistency with our model assumptions, we enforce $\lambda^F_j \geq \lambda^C_j$ through resampling when necessary. This configuration allows us to clearly observe the impact of temptation on consumption choices and enrichment.

To investigate how the platform's knowledge affects recommender performance, we simulate two levels of information availability:

\begin{enumerate}[leftmargin=*]
    \item \textbf{Perfect information:} The platform has complete knowledge of the enrichment and temptation of all on-platform items and off-platform outside options for each user, the probability distribution governing the availability of outside options, and the user-specific feedback and choice parameters. The only unknown in this scenario is the specific outside option available in each round.
    \item \textbf{Partial information:} The platform only has access to historical data and information derived from the additional assumptions described in section \ref{estimation}.
\end{enumerate}

This setup yields $3\times 2 = 6$ distinct scenarios. For each scenario, we simulate consumption over $75$ rounds. The first $25$ rounds serve as \textit{warm-up rounds}, where users receive recommendations of $15$ randomly chosen items. This warm-up period mitigates the impact of a \textit{cold start} on recommender performance. In the subsequent $50$ rounds, users receive recommendations of $15$ items selected by a specific recommendation algorithm.

We evaluate our proposed recommendation strategy against four baseline algorithms:

\begin{enumerate}[leftmargin=*]
\item \textbf{Purely enrichment-based recommendation:} Recommends items solely based on enrichment $u_j(i)$.
\item \textbf{Purely temptation-based recommendation:} Recommends items solely based on temptation $v_j(i)$.
\item \textbf{Ratings-based recommendation:} Generates personalized recommendations based on user-provided ratings, recommending items according to rating $r_j(i)$.
\item \textbf{Click-based recommendation:} Generates personalized recommendations using click data (i.e., recommendations shown to a user and their choices), recommending items based on the choice score $\lambda^{C}_j u_j(i) + (1 - \lambda^{C}_j)v_j(i)$.
\end{enumerate}

To ensure robustness of our results, we repeat our simulation $5$ times for each of the five recommendation algorithms in each scenario, and report the aggregate outcomes.

\subsection{Results}
To assess the efficacy of recommendation algorithms in ensuring user enrichment from consumption, we introduce the metric of \textbf{\textit{overall individual enrichment}}. This measure quantifies the total enrichment an individual user receives, on average, from 50 rounds of consumption across both on-platform and off-platform options, following the initial warm-up period. This comprehensive metric allows us to evaluate how well each algorithm balances the immediate appeal of content with its inherent value to a user. 

\begin{figure*}[t]
    \centering
    \includegraphics[width=\linewidth]{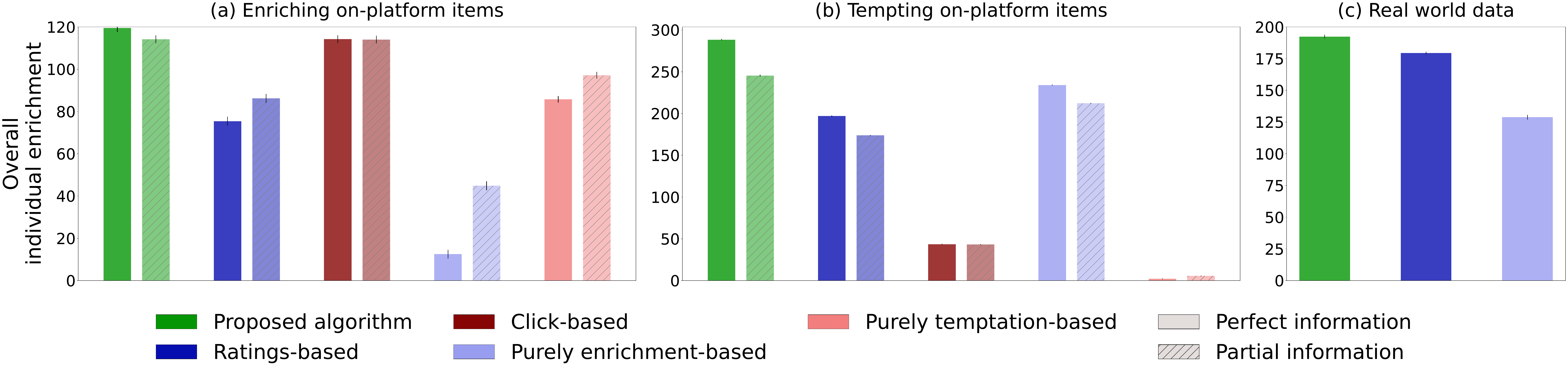}
    \caption{Overall individual enrichment provided by recommendation algorithms. (a) shows algorithm performance for enriching on-platform items, while (b) shows performance for tempting on-platform items. The striped bars show performance when the platform has partial information only. (c) shows algorithm performance in simulations based on real world data from MovieLens. Our proposed algorithm outperforms baselines in all scenarios.}
    \label{fig:all_performance}
\end{figure*}

For brevity, we omit results for the scenario where on and off-platform items have the same enrichment and temptation on average, and report results for four distinct scenarios. \textbf{Figure \ref{fig:all_performance}(a)} and \textbf{\ref{fig:all_performance}(b)} illustrates the performance of all five recommendation algorithms, including our proposed strategy, across four distinct scenarios. These scenarios represent different combinations of platform knowledge and outside option characteristics, providing a holistic view of algorithm performance under varied conditions.

\textbf{Figure \ref{fig:all_performance}(a)} shows the overall individual enrichment when on-platform items offer higher enrichment and lower temptation compared to outside options. Traditional recommenders falter here by potentially suggesting low-temptation items that fail to prevent users from choosing low-enrichment outside options. Consider the example where a user's outside option is online gambling. Recommending an informational video on large language models in this case is unlikely to be effective. Our algorithm addresses this challenge by recommending high-temptation on-platform items that still provide better enrichment than the outside option. In the gambling example, our algorithm might suggest a pop science video such as MythBusters, keeping the user engaged on the platform and away from gambling while also providing enrichment. Notably, click-based and purely temptation-based algorithms outperform ratings-based or purely enrichment-based ones in this scenario as well, simply by keeping users engaged on the platform. In the gambling example, these algorithms are likely recommending cat videos to users, which do not have much enrichment but keeps the users away from gambling regardless.

Conversely, \textbf{figure \ref{fig:all_performance}(b)} depicts the overall individual enrichment when on-platform items offer lower enrichment and higher temptation compared to outside options. In this scenario, traditional recommenders (e.g., ratings-based or click-based) primarily diminish user enrichment by suggesting on-platform items with higher temptation, thereby deterring users from selecting more enriching outside options. Consider the example where a video streaming platform might keep a student from exam preparation by recommending cat videos. Purely enrichment-based recommendations also fail to address this issue, as the outside option could surpass all on-platform items in enrichment while falling short in choice score. Our proposed algorithm, recognizing this complex dynamic, recommends items that guide users towards the optimal enrichment, potentially outside the platform. In the student example, our algorithm might suggest an informational video on large language models, likely prompting the student to return to exam preparation.

\textbf{Figures \ref{fig:all_performance}(a)} and \textbf{\ref{fig:all_performance}(b)} also demonstrate the overall individual enrichment provided by various recommendation algorithms when the platform lacks perfect world knowledge, for different types of outside options. In these scenarios, algorithms estimate necessary information from available historical data, testing their robustness and adaptability. Remarkably, even when estimation is required, our algorithm consistently outperforms all four baselines across different on-platform item scenarios. This consistent superior performance validates the effectiveness of our proposed estimation framework (section \ref{estimation}) and demonstrates the algorithm's resilience to imperfect information. It suggests that our approach can maintain its advantages even in more realistic, information-limited environments, making it a promising candidate for practical implementation in real-world recommendation systems.

\setlength{\textfloatsep}{-100pt}

\section{Evidence from Real World Data}\label{real_data}
To demonstrate the effectiveness of our algorithm in a real-world scenario, we utilize the MovieLens 32M dataset \cite{MovieLens}\footnote{https://grouplens.org/datasets/movielens/}. As of May 2024, this dataset encompasses 32 million movie ratings from approximately 201,000 users across 87,500 movies, providing a rich source of explicit user feedback.

\paragraph{\textbf{Generating click data:}} While the MovieLens dataset provides a rich source of explicit user ratings, it lacks crucial click data, which includes information on what movies were recommended and which ones were chosen. This click data is essential for our algorithm to function effectively. To overcome this limitation, we devise a method to simulate click data using the available timestamp information associated with each rating in the dataset. Our simulation process begins by randomly sampling a set $\mathcal{J}$ of $m$ users and a set $\mathcal{I}$ of $n$ movies from the dataset. For each sampled user, we further sample 25 of their ratings, with each sampled rating corresponding to a single round of movie consumption by the user. This approach allows us to create user interactions that mimic real-world engagement with a movie recommendation platform. We perform this sampling and subsequent computation $5$ times and report the aggregate outcome to ensure robustness.

To incorporate the concept of outside options in our model, we treat ratings for movies that are not part of our sampled set $\mathcal{I}$ as instances of the user consuming an outside option. Using the timestamp information associated with each rating, we reconstruct a chronological sequence of user interactions. For each consumption round, we generate recommendations based on a ratings-based recommender system, considering all ratings provided by the sampled users for the sampled movies up to the time of the current round. We acknowledge that this approach may not perfectly replicate the original data generation process of MovieLens. However, given the widespread use of ratings-based recommenders in real-world applications, we believe it provides a reasonably constructed sandbox environment. This simulated environment allows us to test and evaluate our proposed algorithm in a setting that closely approximates real-world conditions while working within the constraints of the available data.

\paragraph{\textbf{Estimating underlying model:}} Given that the MovieLens website is not a content platform itself and is primarily geared towards collecting explicit user feedback (i.e., rating) from users, we think it is reasonable to assume that the ratings on MovieLens are not affected by temptation. Consequently, we assume that $\lambda^F_j = 1 \forall j\in \mathcal{J}$. This implies that the ratings provided by users represent the enrichment of items. We also assume that the choice scores of ``outside options'' (movies that were rated by the user during the sampled consumption rounds but are not in the set of sampled movies) come from a normal distribution.

With these assumptions in hand, we use our estimation framework from section \ref{estimation} to compute the temptation of all sampled movies for all sampled users, the choice parameter of each sampled user, and the distribution parameters for the outside options. We treat these estimated parameters as perfect information about the world the users and movies are in, which then allows us to observe the effects of recommendation algorithms on user consumption of movies. As section \ref{estimation} demonstrates, our estimation framework does well in emulating the perfect information scenario from historical data, which validates our approach to use estimated model parameters in constructing our sandbox.

\subsection{Results}
Using the estimated user-movie interaction model, we simulate 50 rounds of movie consumption. In each round, users are presented with a set of recommended movies from a specific algorithm and make choices based on estimated choice scores. We evaluate the overall individual enrichment for an average user across these 50 rounds, comparing three recommendation algorithms: our proposed algorithm, the ratings-based recommender (equivalent to a purely enrichment-based recommender given $\lambda^F_j = 1$), and the click-based recommender. This simulation allows us to assess the real-world effectiveness of our approach compared to traditional recommendation methods in a controlled yet realistic setting.

\textbf{Figure \ref{fig:all_performance}(c)} shows the average overall individual enrichment received from the $50$ rounds of simulated consumption. As seen here, our proposed algorithm outperforms both the ratings-based and the click-based recommenders, proving its effectiveness even when applied to consumption of real items by real users.

\section{Recommendation Design Philosophy}

Traditional recommender systems have predominantly adopted a predictive role in understanding and guiding user behavior on online platforms, often leaving the responsibility of self-control and consumer well-being entirely to users themselves. This approach, while seemingly neutral, can inadvertently lead to suboptimal outcomes for users, particularly when confronted with the reality of temptation in content consumption. For example, as we have remarked before in section \ref{recommender_design}, offering users more ``freedom'' of choice cannot hurt in a world without temptation, because users could simply not choose the additional recommended item(s). However, in the presence of temptation, the additional recommended item(s) could have high temptation and low enrichment, hindering users from choosing an item that would otherwise offer better enrichment.  

The act of recommending tempting but unenriching content can be conceptualized as creating a form of harm: it generates desires that are costly to fulfill yet unsatisfying to leave unmet. This reveals a subtle challenge for recommender systems, as eliciting engagement signals from users may be more costly than previously recognized. Ideally, platforms seek to learn user preferences by recommending diverse sets of items~\cite{Peng2023}. However, larger option pools are more likely to contain highly tempting options that deter users from selecting more enriching alternatives. This creates a fundamental trade-off for recommenders between maximizing user enrichment and gathering information about user preferences.

Traditional recommenders do not account for the complex dynamics of temptation and enrichment, leaving everything up to the users and therefore providing little help in overcoming temptation. Our proposed recommendation algorithm diverges from this predictive stance, instead adopting a more prescriptive role that actively seeks to guide users towards consumption with higher enrichment. This shift in approach is similar in spirit to the concept of {\it libertarian paternalism}, as conceptualized by Sunstein and Thaler \cite{Thaler2003}, which posits that private and public institutions should guide people towards better versions of themselves while still respecting their freedom of choice.  Crucially, note that the role of our proposed recommender is not a ``pro-social'' one \textemdash we are not imposing any moral constraints beyond users' underlying preferences or desires via our recommender. Instead, we merely uncover user preferences or desires otherwise obscured by temptation, and provide better navigation towards them.  Any system must wrestle with the choice of whether to optimize for user's perceived enrichment or the user's perceived desires under the revealed preference assumption. This is not really a ``paternalistic" a choice of any kind, rather, in the conception of the dual self, an choice of which ``self'' to serve.  Perhaps users could, at the beginning, be offered a choice between our recommendation system and an engagement-based approach.  In the dual self conception of temptation, the users would always choose our system because it will better help to achieve the goals the ``self" that is doing the choosing.  

\begin{figure}
    \centering
    \includegraphics[width=\linewidth]{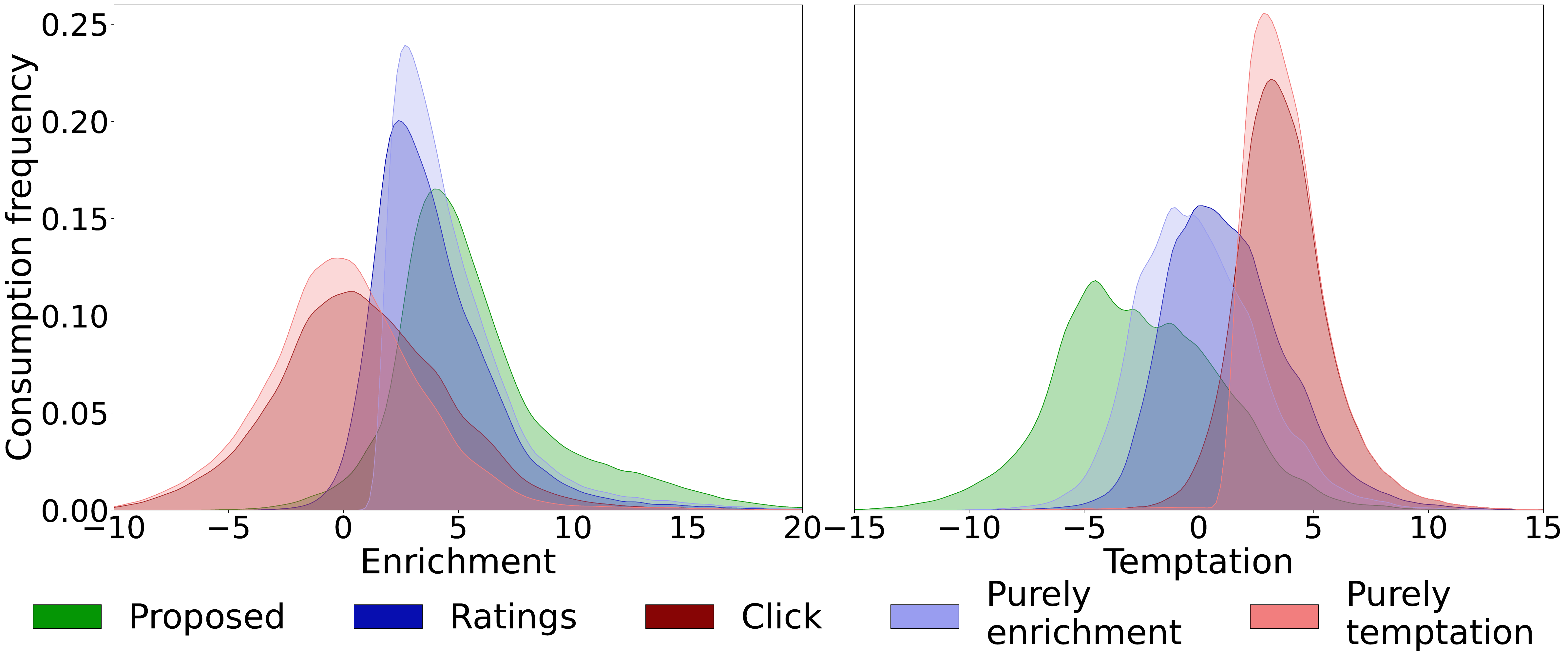}
    \caption{Frequency of consumption against enrichment and temptation. In the presence of our algorithm, user consumption on the platform is skewed towards higher enrichment and lower temptation, incentivizing creation of better content.}
    \label{fig:freq}
\end{figure}

The impact of recommender systems extends beyond individual user experiences to influence broader platform dynamics, particularly on the supply side. Content creators, observing consumption patterns heavily influenced by recommendation algorithms, make decisions about what kind of content to produce, balancing between enrichment and temptation. Our research demonstrates the significant role that recommender systems play in shaping these decisions. As illustrated in \textbf{Figure \ref{fig:freq}}, which shows the frequency of consumption across varying levels of enrichment and temptation for different recommendation algorithms, our proposed algorithm shifts user consumption on the platform towards items with higher enrichment and lower temptation. This shift has far-reaching implications for content creation. By skewing consumption towards more enriching content, our algorithm creates an incentive structure that encourages content creators to focus on producing high-quality, enriching content rather than relying on temptation-driven engagement. Over time, this can lead to an overall improvement in the quality of available content on the platform. This virtuous cycle of improved recommendations leading to better content creation demonstrates the potential of thoughtfully designed recommender systems to positively transform the entire ecosystem of online platforms, benefiting both users and content creators alike.

\setlength{\textfloatsep}{5pt}

\section{Conclusion and Future Work}

In this paper, we addressed a key limitation of classic recommender systems: the assumption that users are enrichment maximizers, despite often succumbing to the temptation of instant gratification. We introduced a novel model of user consumption behavior that disentangles enrichment and temptation and accounts for the existence of outside options. We highlight the challenge in recognizing what to optimize for when recommending content, proposed the appropriate objective for enrichment maximization, and developed the optimal strategy under perfect information. Our approach identifies the information about user experience outside the platform necessary beyond historical consumption data on the platform, and outlines a basic estimation framework using minimal assumptions. Through synthetic simulations and simulations based on real-world data from MovieLens, we demonstrated that our proposed algorithm has the potential to outperform traditional recommenders, providing users with superior enrichment from their consumption both on and off the platform. We note that if our estimation procedure is improved, this may lead to additional performance gains for our proposed recommendation strategy.

Our work opens several promising avenues for future research. First, improving the accuracy of our estimation framework could bring the algorithm's performance closer to its theoretical optimum under perfect information. Second, further investigation into data collection methods, specifically eliciting information about outside options, and the validity of our assumptions would be invaluable for the practical implementation of our algorithm. Extending the model to consider multiple competing platforms and the dynamics of user behavior in a multi-platform environment represents another exciting direction. This could involve exploring how users navigate between platforms and how their preferences evolve in response to recommendations across different platforms.

In conclusion, our work represents a significant step towards more responsible and user-centric recommender systems that account for the complex nature of human decision-making. As we continue to refine and expand upon these ideas, we move closer to a future where digital platforms not only engage users but genuinely enrich their lives.

\bibliographystyle{ACM-Reference-Format}
\balance
\bibliography{ref_recsys_2025}

\appendix
\include{supp_recsys_2025}

\end{document}

%% file: supp_recsys_2025.tex
\balance

\section{Proof of Theorem 1}

\begin{theorem}\label{main_theorem}
    When the platform has perfect knowledge of everything except the exact availability of outside options, the optimal recommendation strategy is \textbf{locally greedy}, i.e. in each round $t$, it recommends the available item that maximizes the expected enrichment a receives from consumption in a single round when the available options are the said item and the outside option.
    
    In other words, the optimal strategy recommends the available item $i^*$ in round $t$ such that
    \[i^* = \operatorname*{argmax}_{i}\mathbb{E}_{o_j(t)}\left[u_j\left(i|o_j(t)\right)\right]\]
\end{theorem}

\begin{proof}
    Note that since users choose what to consume via a deterministic process, recommending a set $S_j$ of items to user $j$ is equivalent to recommending item $i_{choice}$, where
    \[i_{choice} = \operatorname*{argmax}_{i\in S_j} \left (\lambda^C_j u_j(i) + \left(1 - \lambda^C_j\right) v_j(i)\right ).\]

    Therefore, every recommendation strategy can be represented as a binary tree, where the depth of the tree represents the number of rounds, each node represents a recommended item, and branching is done based on whether the recommended item was chosen or not. To prove our claim, we induct on the number of rounds, $T$. The base case, $T = 1$, is trivial, since our recommendation objective in this case becomes the expected enrichment from one round. Assume that the claim holds for some $T$. 
    
    Let $\mathcal{A}$ be the optimal recommendation strategy for $T + 1$ rounds, and let $\mathcal{G}$ be the locally greedy one. Let $\mathcal{T}(\mathcal{A})$ and $\mathcal{T}(\mathcal{G})$ be the corresponding tree representations of these strategies. Let $A\rightarrow_{Y}B$ denote an edge in these trees where item $A$ was recommended and consumed, and then item $B$ was recommended. Define $A\rightarrow_{N}B$ analogously. Note that we can write any path in these trees as a collection of edges from item to item. Let $i_k$ denote the $k$th new item the locally greedy strategy would recommend. 
    
    By the induction hypothesis, after the first round of consumption, $\mathcal{A}$ must be locally greedy. If $\mathcal{A}$ recommends item $i_1$ in the first round (same as the locally greedy strategy), then we are done. So, assume that during the first round, $\mathcal{A}$ recommends item $i_k$ for some $k \neq 1$. 
    
    Our plan is as follows: each path in $\mathcal{T}(\mathcal{A})$ corresponds to a leaf.  We will create a matching between the paths/leaves of $\mathcal{A}$ (and sometimes pairs of such leaves/paths) and those of $\mathcal{G}$.    We will show that each path/leave (pair) in $\mathcal{G}$ happens with exactly the same probability as the corresponding path/leave (pair) in $\mathcal{A}$, but its reward is at least as much.

    Note that since $\mathcal{A}$ is locally greedy after round 1, any path $p$ in $\mathcal{T}(\mathcal{A})$ has the form $i_k\rightarrow p' \rightarrow i_\ell$, where the subpath $p'$ begins with $i_1$ and proceeds in order. Consider such a path $p$. We perform a case analysis:

    \begin{itemize}
        \item \textbf{Case 1:} $p = i_k \rightarrow_Y p' \rightarrow i_\ell$ and $\ell > k$.  In this case $i_k$ does not appear in $p'$ because it was already selected.  We construct $p^*$ by starting with $p'$ and replacing the edge $i_{k-1}\rightarrow_Y i_{k+1}$  with the edges $i_{k-1}\rightarrow_Y i_{k} \rightarrow_Y i_{k+1}$.  Note that $p^*$ is a path in $\mathcal{T}(\mathcal{G})$ which occurs with exactly the same probability as $p$ because the order of the edges is just permuted.   
        \item \textbf{Case 2:} $p = i_k \rightarrow_N p' \rightarrow i_\ell$ and $\ell \geq k$.  In this case $i_k$ must appear in $p'$.  We construct $p^*$ by starting with $p'$ and replacing the edge $i_{k-1}\rightarrow_Y i_{k}$ with the edges $i_{k-1}\rightarrow_Y i_{k} \rightarrow_N i_{k}$.  Again, note that $p^*$ is a path in $\mathcal{T}(\mathcal{G})$ which occurs with exactly the same probability as $p$ because the order of the edges is just permuted.  
         \item \textbf{Case 3:} $\ell < k$.
         In this case, $p'$ does not contain $k$ and we combine both the analysis for the two paths $p = i_k\rightarrow_Y p' \rightarrow i_\ell$ and $\hat{p} = i_k\rightarrow_N p' \rightarrow i_\ell$.  We construct $p^*$ by starting with $p'$ and adding the edge $i_{\ell} \rightarrow_Y i_{\ell+1}$,  and we construct $\hat{p}^*$ by starting with $p'$ and adding the edge $i_{\ell} \rightarrow_N i_{\ell}$.  Note that the pair $p$ and $\hat{p}$ occur with the same probability that $p'$ does.  Similarly, $p^*$ and $\hat{p}^*$ also occur with the same probability that that $p'$ does.  However, in the former pair, there is one chance for the item $i_k$ to be selected, and in the later there is one (addition) chance for item $i_{\ell}$ to be selected.  Because $\ell < k$ in the greedy ordering, we know that the expected reward for $\ell$ is greater than or equal to the expected reward for $k$.  Apart from this difference, the rewards are equal (because both branches select exactly the same items: $i_1$, $i_2$, \ldots, $i_{\ell-1}$.
    \end{itemize}
    Finally, we note that this is a bijection because both sets have the same size and the inverse is defined.  To map from the tree  $\mathcal{T}(\mathcal{G})$ back to  $\mathcal{T}(\mathcal{A})$ there are two cases.  The first, $i_k$ occurs at some time, and then we simply move the first edge with $i_k$ to the front of the path.  In the second, $i_k$ never occurs and we move add $i_k$ to the beginning of the path and remove the final edge.  
    
    This shows that the recommendation strategy $\mathcal{A}$, which we assumed to be optimal for $T + 1$ rounds, can do at most as well as the locally greedy strategy $\mathcal{G}$. This completes the inductive hypothesis and our proof.
\end{proof}

\section{Simulation Parameter Selection}
As mentioned in section \ref{simulation}, we draw the $d$-dimensional enrichment and temptation space representation vectors for users and items from multivariate normal distributions ($d = 3$).

The enrichment and temptation space representations $\mathbf{a}_j$ and $\mathbf{b}_j$ of users $j$ are drawn in a way such that:
\begin{enumerate}
    \item The first components are $1$: for all user $j = 1, \ldots, m$, 
    \[\mathbf{a}_{j1} = \mathbf{b}_{j1} = 1\]
    \item The remaining components are each drawn from a normal distribution: for all user $j = 1, \ldots, m$, 
    \begin{align*}
        &\mathbf{a}_{j\ell} \sim \mathcal{N}(0, 2.5)\\
        &\mathbf{b}_{j\ell} \sim \mathcal{N}(0, 2.5)
    \end{align*}
    \item The corresponding components of the enrichment and temptation representations are strongly anti-correlated: for all user $j = 1, \ldots, m$, 
    \[\mathrm{Cov}\left(\mathbf{a}_{j\ell}, \mathbf{b}_{j\ell}\right) = -1\].
\end{enumerate}

User-specific choice and feedback parameters from Beta distributions: for all user $j = 1, \ldots, m$,
\begin{align*}
    \lambda^C_j &\sim \mathrm{Beta}\left(12.5, 37.5\right)\\
    \lambda^F_j &\sim \mathrm{Beta}\left(37.5, 12.5\right)
\end{align*}

\begin{figure*}[t]
    \centering
    \includegraphics[width=\linewidth]{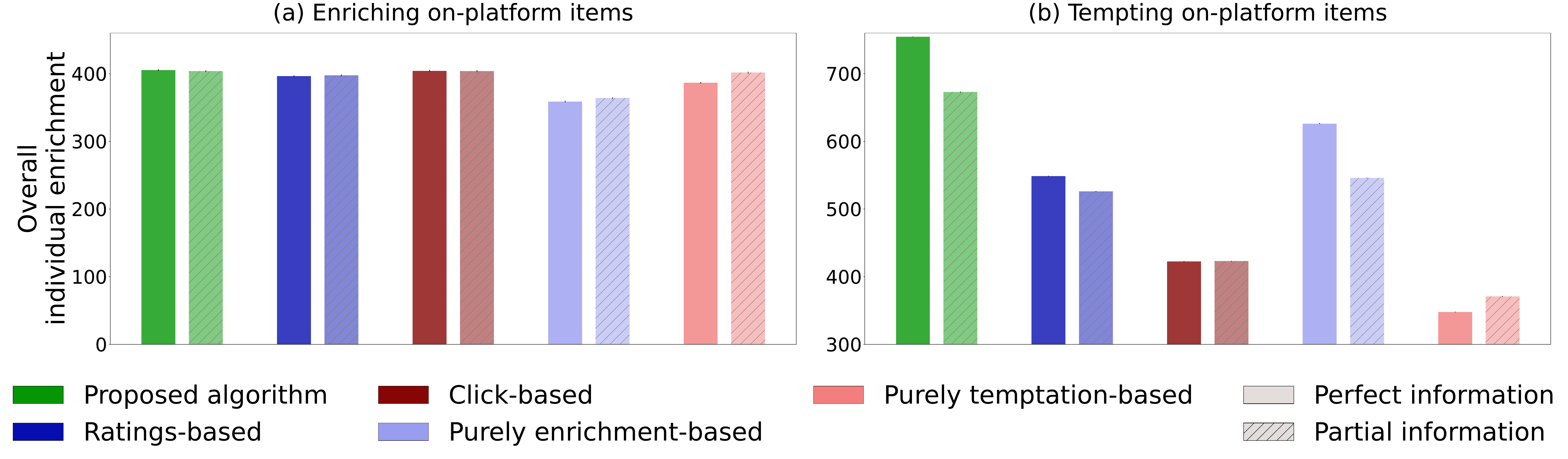}
    \caption{Overall individual enrichment provided by recommendation algorithms when enrichment and temptation of on-platform items have skewed, fat-tailed distributions. (a) shows algorithm performance for enriching on-platform items, while (b) shows performance for tempting on-platform items. The striped bars show performance when the platform has partial information only. Our proposed algorithm maintains superior performance.}
    \label{fig:robustness}
\end{figure*}

In particular, to ensure that $\lambda^C_j \leq \lambda^F_j$ for each user $j$, we resample $\lambda^C_j$ and $\lambda^F_j$ whenever necessary.

The enrichment and temptation space representations of on-platform items are drawn in a way such that:
\begin{enumerate}
    \item The first components (representing universal enrichment and temptation) are drawn from a a normal distribution and are strongly anti-correlated: for all items $i = 1, \ldots, n$,
    \begin{align*}
        &\mathbf{x}_{i1} \sim \mathcal{N}(10, 10)\\
        &\mathbf{y}_{i1} \sim \mathcal{N}(0, 10)\\
        &\mathrm{Cov}\left(\mathbf{x}_{i1}, \mathbf{y}_{i1}\right) = -1
    \end{align*}
    \item The remaining components are each drawn from a different normal distribution: for all items $i = 1, \ldots, n$ and for all remaining dimensions $\ell = 2, \ldots, d$,
    \begin{align*}
        &\mathbf{x}_{i\ell} \sim \mathcal{N}(0, 1)\\
        &\mathbf{y}_{i\ell} \sim \mathcal{N}(0, 1)
    \end{align*}
\end{enumerate}

The enrichment and temptation space representations of off-platform outside options are drawn in a way such that
\begin{enumerate}
    \item The first components (representing universal enrichment and temptation) are drawn from a a normal distribution and are strongly anti-correlated: for all outside options $o = o^1, \ldots, o^K$,
    \begin{align*}
        &\mathbf{x}_{o1} \sim \mathcal{N}(\mu_x, 10)\\
        &\mathbf{y}_{o1} \sim \mathcal{N}(\mu_y, 10)\\
        &\mathrm{Cov}\left(\mathbf{x}_{o1}, \mathbf{y}_{o1}\right) = -1
    \end{align*}
    \textbf{where $\mu_x$ and $\mu_y$ depend on the specific scenario.}
    \item The remaining components are each drawn from a different normal distribution: for all outside options $o = o^1, \ldots, o^K$ and for all remaining dimensions $\ell = 2, \ldots, d$,
    \begin{align*}
        &\mathbf{x}_{o\ell} \sim \mathcal{N}(0, 1)\\
        &\mathbf{y}_{o\ell} \sim \mathcal{N}(0, 1)
    \end{align*}
\end{enumerate}

As mentioned in section 5, we simulate three distinct scenarios to vary the relative enrichment and temptation of on-platform items and off-platform outside options:
\begin{enumerate}
    \item \textbf{Enriching on-platform items:} We set $\mu_x = -5$ and $\mu_y = 35/3$ so that on-platform items have higher enrichment and lower temptation than outside options on average.

    \item \textbf{Tempting on-platform items:} We set $\mu_x = 15$ and $\mu_y = -10$ so that on-platform items have lower enrichment and higher temptation than outside options on average.

    \item \textbf{Similar on and off-platform items:} We set $\mu_x = 10$ and $\mu_y = 0$ so that on-platform items have the same enrichment and temptation as outside options on average.
\end{enumerate}

These specifications for the outside option vector representations allow us to observe the effect of recommendation algorithms on user enrichment in the presence of different kinds of outside options more clearly. In favor of saving space, we report the results from only the first two scenarios in our paper; however, our results hold for the third scenario as well.

\section{Robustness Tests}

In realistic scenarios, enrichment and temptation of items for a given user may not be normally distributed. To ensure that our estimation framework is robust to the distributions of these quantities, we reran our simulated experiments by drawing the first components of representations $x_{i1}, y_{i1}$ of each on-platform item $i$ (representing universal enrichment and temptation) from skewed, fat-tailed Johnson's $S_u$ distributions. Specifically,
\begin{align*}
    x_{i1}&\sim Johnson-S_U\left(3.25,1,12.3520,0.3933\right)\\
    y_{i1}&\sim Johnson-S_U\left(3.25,1,2.3520,0.3933\right)
\end{align*}
We kept the distribution of enrichment and temptation of outside options identical to our previous experiments.

Figure \ref{fig:robustness} shows the performance of our algorithm against four baseline algorithms in four possible scenarios described in section 5. As the figure shows, our algorithm maintains its superior performance, demonstrating its robustness to distributions of enrichment and temptation.